\documentclass[a4paper,11pt]{article}
\usepackage{pos}

\title{An overview of the JEM-EUSO program and results}
 \ShortTitle{The JEM-EUSO Program}

\author*[a]{M.~Bertaina}

\affiliation[a]{Department of Physics, University of Torino \& INFN Torino,\\
  Via P. Giuria 1, Torino, Italy}

\forColl{JEM-EUSO} 

\emailAdd{bertaina@to.infn.it}

\abstract{

The field of UHECRs (Ultra-High energy cosmic Rays) and the understanding of particle acceleration in the cosmos, 
as a key ingredient 
to the behaviour of the most powerful sources in the universe, is of outmost importance for 
astroparticle physics as well as for fundamental physics and will improve our general understanding 
of the universe. The current main goals are to identify sources of UHECRs and their composition. 
For this, increased statistics is required. A space-based detector for 
UHECR research has the advantage of a very large exposure and a uniform coverage of the celestial sphere. 
The aim of the JEM-EUSO program~\cite{EUSO-Program} is to bring the study of UHECRs to space. 
The principle of observation is based on the detection of UV light emitted by isotropic ﬂuorescence 
of atmospheric nitrogen excited by the Extensive Air Showers (EAS) in the Earth’s atmosphere and 
forward-beamed Cherenkov radiation reﬂected from the Earth’s surface or dense cloud tops. 
In addition to the prime objective of UHECR studies, JEM-EUSO will do several secondary studies due 
to the instruments' unique capacity of detecting very weak UV-signals with extreme time-resolution 
around 1 $\mu$s: meteors, Transient Luminous Events (TLE), bioluminescence, maps of human generated UV-light, 
searches for Strange Quark Matter (SQM) and high-energy neutrinos, and more.
The JEM-EUSO program includes several missions from ground (EUSO-TA~\cite{eusota}), from stratospheric balloons
(EUSO-Balloon~\cite{eusobal}, EUSO-SPB1~\cite{spb1}, EUSO-SPB2~\cite{spb2}),
and from space (TUS~\cite{tus}, Mini-EUSO~\cite{minieuso}) employing fluorescence
detectors to demonstrate the UHECR observation from space and
prepare the large size missions K-EUSO~\cite{keuso} and POEMMA~\cite{poemma}.
A review of the current status of the program, the key results 
obtained so far by the different projects, and the perspectives for the near future are presented.

}

\FullConference{37$^{\rm{th}}$ International Cosmic Ray Conference (ICRC 2021)\\
		July 12th -- 23rd, 2021\\
		Online -- Berlin, Germany}

\begin{document}
\maketitle

\section{The ground-based telescope EUSO-TA}
\label{eusota}
\begin{figure}[t]
\begin{center}
\includegraphics[width=0.9\textwidth]{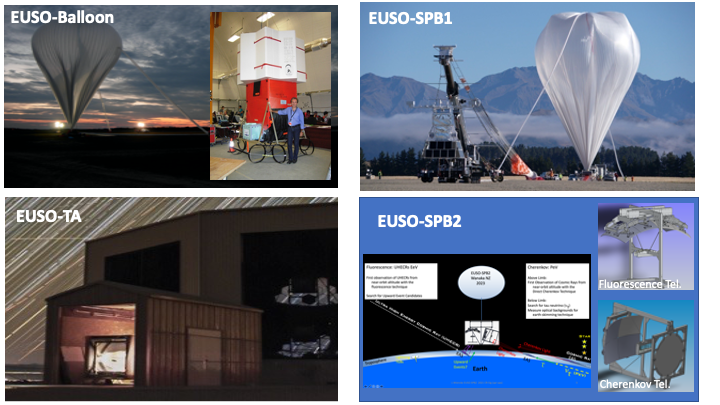}
\end{center}
\caption{EUSO-TA and stratospheric balloon missions of the JEM-EUSO program.
See text for details.}
\label{fig:detector}
\end{figure}

EUSO-TA is a ground-based telescope, installed at the TA site in Black Rock
Mesa, Utah, USA (see Fig.~\ref{fig:detector}). This is the first detector to successfully use a Fresnel lens
based optical system and Multi-Anode Photomultipliers (MAPMT). EUSO-Balloon,
EUSO-SPB1 and Mini-EUSO adopted the same observational approach.
Each MAPMT has 64 channels, and they are grouped in blocks of 2$\times$2 to
form one Elementary Cell (EC). Nine ECs form one Photo-Detector Module (PDM)
which has, therefore, 2304 channels encompassing a $10.6^{\circ} \times
10.6^{\circ}$ field of view (FoV) for the detection of UHECRs.
The telescope is located in front of one of the
fluorescence detectors of the TA experiment. Since its first operation in 2015, a few campaigns of
joint observations allowed EUSO-TA to detect 9 UHECR events in 120 hours of data taking, as well as
a few meteors. The limiting magnitude of 5.5 on summed frames ($\sim$3 ms) has been established. 
The detector is also being calibrated using stars in the
FoV, as discussed in~\cite{plebaniak_icrc19}.
These observations provide important data to optimise the detector technology
in view of the upcoming space-based missions and to determine the sensitivity of
the experiment (see~\cite{bisconti_icrc19}). 
The future upgrades of the detector include a new acquisition system based on
Zynq board
and the implementation of a self-triggering system. 
Simulations indicate that the new configuration should be able to have a self-trigger rate 
at the level of $\sim$1 UHECR 
per 24 hours of accumulated measurement.

\section{The balloon program: EUSO-Balloon, EUSO-SPB1 \& EUSO-SPB2}
\label{sec:euso-bal}
The JEM-EUSO program includes three stratospheric balloon missions
with increasing level of performance and upgraded designs (see Fig.~\ref{fig:detector}). In addition to
demonstrating the capabilities of the JEM-EUSO instruments to detect and
reconstruct EAS from the edge of space, they also give access to direct
measurement of the UV nightglow emission and artificial UV contributions above
ground and oceans, which are important information to optimise the design of
the space-based missions.
Two balloon flights have been performed so far: EUSO-Balloon (Canada, 1 night)
and EUSO-SPB1 (New Zealand,
12 nights). A third one (EUSO-SPB2) is scheduled to fly
in 2023.

EUSO-Balloon~\cite{eusobal} was launched by CNES from
the Timmins base
in Ontario (Canada) on the moonless night of August 25, 2014. 
The telescope configuration was very similar to EUSO-TA and the subsequent EUSO-SPB1.
After
reaching the floating altitude of $\sim$38 km, EUSO-Balloon imaged the UV
intensity with a spatial and temporal resolutions of 130 m and
2.5 $\mu$s, respectively, in the wavelength range 290 - 430 nm for more than 5 hours before
descending to ground.
The full FoV in nadir mode was $\sim11^{\circ}$.
During 2.5 hours of EUSO-Balloon flight, a
helicopter circled under the balloon operating UV flashers and
a UV laser to simulate the optical signals from UHECRs, to calibrate the
apparatus, and to characterise the optical atmospheric conditions. Data collected
by EUSO-Balloon have been analysed
to infer different information among which the response of the detector to the
UV flasher and laser events, and the UV radiance from the Earth atmosphere and
ground in different conditions: clear and cloudy atmosphere, forests,
lakes, as well as city lights.
This is relevant for a JEM-EUSO-like mission
as it is one of the key parameters to estimate the exposure curve as a
function of energy \cite{exposure}.
The results of this analysis are reported in~\cite{kenji-uv}
and are in the band of previous measurements confirming a good
understanding of the detector performance also in this respect.
The helicopter events proved to be extremely useful to understand the system's
performance and to
test the capability of EUSO-Balloon to detect and reconstruct signals similar
to EAS~\cite{eser-jinst}.
The data collected by EUSO-Balloon were used together with those collected by
EUSO-TA to define an internal
trigger logic~\cite{trigger} that was implemented on the EUSO-SPB1 flight.

EUSO-SPB1~\cite{spb1}
was launched on April 25, 2017 from Wanaka, New Zealand, as a mission of opportunity
on a NASA SPB test flight planned to circle the southern
hemisphere. 
The telescope was an upgraded version of that used in the EUSO-Balloon
mission. An autonomous
internal trigger was implemented according to~\cite{trigger} to detect UHECRs.
Prior to flight, in October 2016, the fully assembled EUSO-SPB1 detector was
tested for a week
at the EUSO-TA site to measure its response and to calibrate it by means of a
portable Ground Laser System (GLS). Observations of Central Laser Facility
(CLF), stars, meteors were performed.
The $\sim$50\% trigger efficiency was reached at laser energies whose
luminosity is equivalent to $\sim$45$^{\circ}$ inclined EAS of E $\sim3 \times
10^{18}$ eV seen from above by a balloon flying at 33 km altitude.
Unfortunately, although the instrument was showing
nominal behaviour and performances, the flight was terminated prematurely in the
Pacific Ocean about 300 km SE of Easter Island after only 12 days
aloft, due to a leak in the carrying balloon.
During flight, $\sim$30 hours of data were collected, the trigger rate was
tipycally a few Hz, which is compliant
with JEM-EUSO requirements~\cite{mario-nim2019}.
A deep analysis of the collected data was performed. Tracks of
CRs directly crossing the detector were recognized. However, no
EAS track has been clearly identified~\cite{diaz_icrc19,vrabel_icrc19}.
Simulations post-flight indicate
that the number of expected events is 0.5 - 1 in the
available data sample including the role of clouds
confirming pre-flight expectations for such a flight
duration~\cite{eser_icrc19,kenji_icrc19}. 

The subsequent step of the JEM-EUSO program development is currently under
realization: EUSO-SPB2~\cite{johannes_icrc21}. It will be equipped with 2
telescopes. One telescope
will be devoted to UHECR measurements using the fluorescence technique.
The Focal Surface (FS)~\cite{osteria_icrc21} will be equipped with 3 PDMs to
increase the UHECR collection power.
EUSO-SPB2 will adopt a Schmidt camera with
a shorter temporal resolution of 1 $\mu$s and a more efficient trigger logic to improve the sensitivity of the 
instrument. 
Simulations indicate that the energy threshold is around 2$\times$10$^{18}$ eV and an
expected detection rate of 0.6 events per night of acquisition~\cite{george_icrc21}.
The FS of the second telescope
is based on SiPMT sensors and a dedicated electronics to detect
the Cherenkov emission in air by UHECR-generated EASs~\cite{mahdi_icrc21}. In
perspective they will test the capability to detect EAS generated by
$\nu_{\tau}$ interacting in the Earth crust~\cite{austin_arxiv}. For this
observation the
detector will be pointing slightly below the limb. The observation above
the limb will allow to study UHECRs through their Cherenkov
emission.  
EUSO-SPB2 is expected to fly in 2023 from Wanaka on a NASA Super
Pressure Balloon.

\section{The first space-based missions of the program: TUS and Mini-EUSO}
\label{sec:tus}
\begin{figure}[t]
\begin{center}
\includegraphics[width=0.9\textwidth]{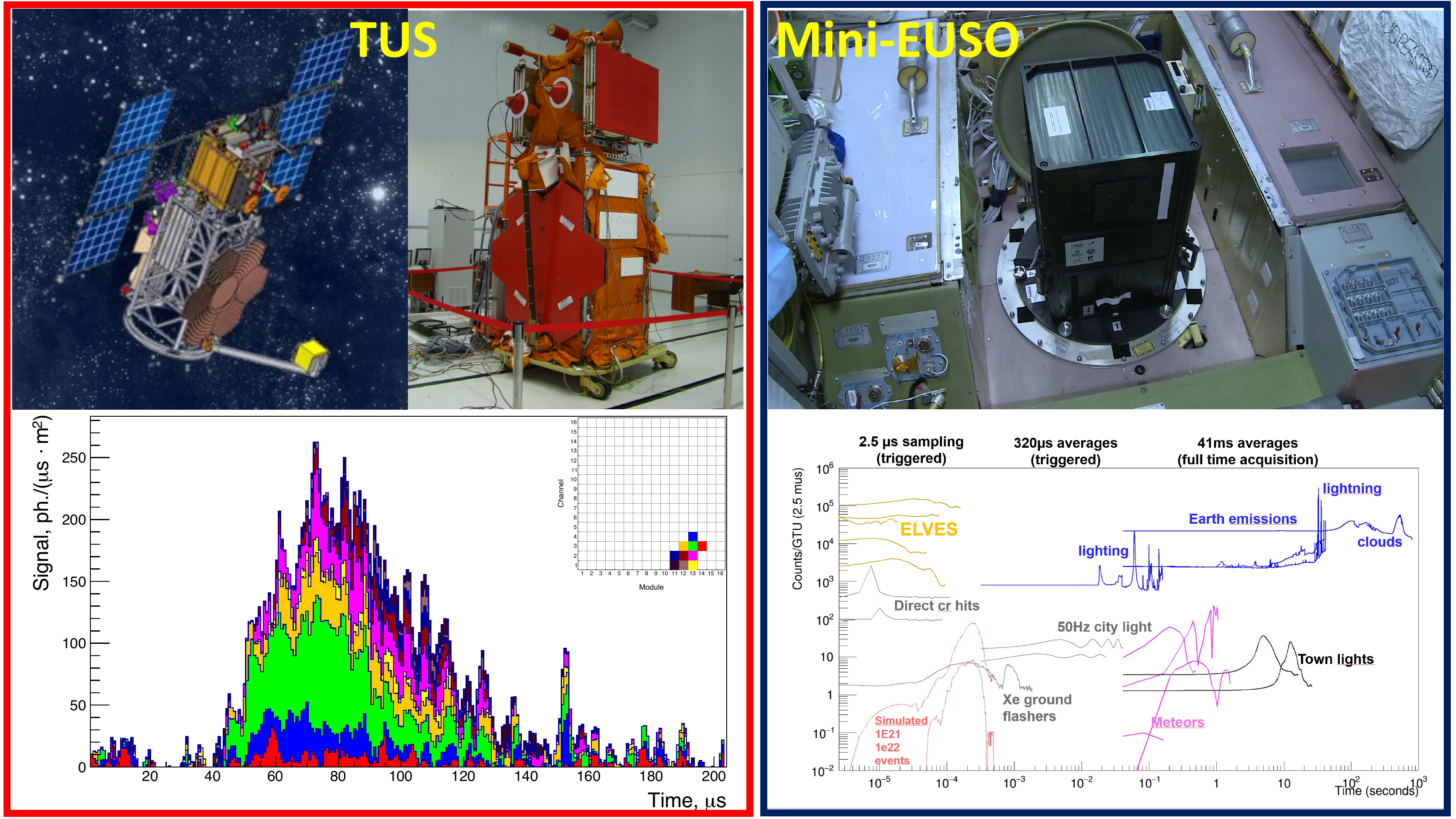}
\end{center}
\caption{TUS and Mini-EUSO telescopes with examples of their observations (see text for details).}
\label{fig:tus-minieuso}
\end{figure}

The Track Ultraviolet Setup (TUS)~\cite{tus} detector
(see Fig.~\ref{fig:tus-minieuso})
was launched on April 28, 2016 as a
part of the scientific payload of the Lomonosov satellite. 
The instrument was actively recording data till November 2017. Different
scientific modes were tested: cosmic ray, lightning and meteor modes.
The satellite has a sun-synchronous orbit with an inclination of 97.3$^{\circ}$,
a period of $\sim$94 min, and a height of 470 - 500 km.
The telescope consists of two main parts: a modular Fresnel mirror-concentrator
with an area of $\sim$2 m$^2$ and 256 PMTs
arranged in a 16$\times$16 photo-receiver matrix located in the focal plane of
the mirror. The pixel's FoV is 10 mrad, which corresponds
to a spatial spot of $\sim$5 km $\times$ 5 km at sea level. 
Thus, the full area observed by TUS at any moment is
$\sim$80 km $\times$ 80 km. TUS is sensitive to the near UV band and in
cosmic ray mode has
a time resolution of 0.8 $\mu$s in a full temporal interval of 256 time steps.
TUS data offer the opportunity to
develop strategies in the analysis and reconstruction of events which will be
essential for future space-based missions. A detailed analysis of TUS exposure and
sensitivity to UHECR is presented in ~\cite{fenu_tus_icrc21}.
During
its operation TUS has detected about 8$\times$10$^4$ events that have been
subject to an offline analysis to select among them those satisfying
basic temporal and spatial criteria of UHECRs. A few events passed this
first screening.
One specific event (see Fig.~\ref{fig:tus-minieuso}) registered in perfect observational conditions was
deeply scrutinized. Its phenomenology and the possible interpretations are
reported in detail in~\cite{tus_jcap}.
Similar searches performed with Mini-EUSO support the interpretation of an anthropogenic
origin of this event. However, the important result of this observation is that
a space-based observatory can detect light pulses with temporal and spatial characteristics of EAS.

While TUS was conceived mainly to prove the observation of UHECRs from space
with a FS-instrumentation similar to ground-based detectors,
Mini-EUSO has been developed in order to test the same FS-instrumentation
foreseen for K-EUSO and POEMMA.
Mini-EUSO~\cite{casolino_icrc21} is a UV telescope launched in August 2019
and installed periodically inside the ISS since October 2019, looking down the Earth from a nadir-facing 
window in the Russian Zvezda module. So far more than 40 sessions of about 12 hours of data taking have been
performed~\cite{cambie_icrc21}. Mini-EUSO maps the Earth in the UV range (290 - 430 nm) with a
spatial resolutions of $\sim$6 km (similar to TUS) and three different temporal resolutions of
2.5 $\mu$s, 320 $\mu$s, and 41 ms, respectively. While the 41 ms time range allows a continuous video-taking, the other
two modes allow acquisitions of 4 packets of 128 GTUs each every 5.24s to catch fast luminous transients (flashes, 
lightnings, etc..). The optical system consists of 2 Fresnel lenses of 25 cm diameter each
with a large FoV of $\sim$44$^{\circ}$. 
Mini-EUSO energy threhsold for UHECRs is well above 10$^{21}$ eV~\cite{fenu_minieuso_icrc21}.
Data carried down to Earth from the ISS allowed to perform
the first analyses showing that Mini-EUSO observes different Earth emissions depending on the surface visible, e.g., 
ground, sea, or clouds as well as slow transients such as meteors (thousands of events have been identified in the data
with absolute magnitude lower than +5 ~\cite{lech_icrc21}).
At shorter times scales, several hundreds of lightings (among them 17 elves~\cite{laura_icrc21}) have been detected and at 
much shorter time
scales, many anthropogenic flashes presumably related to airport lights or other flashing tower lights have been acquired. 
Thanks to the
Mini-EUSO capability to record consecutive events, this class of events is clearly identified as they repeat themselves 
periodically~\cite{matteo_icrc21} and their location can be easily recognized thanks to the continuous data taking in slow mode (41 ms time 
frames). Some of them have characteristics similar to the TUS event previously described. So far, only one event without
repetition and above ocean has been detected. However, a UHECR origin of this event has been excluded based on the light
profile associated to the duration and image on the FS. Preliminary analysis of the UV maps indicate that Mini-EUSO 
often detects the presence of clouds~\cite{alessio_icrc21} and that the UV light intensity recorded by 
Mini-EUSO is within the range of values considered in the estimation of JEM-EUSO exposure~\cite{kenji_icrc21}. 

\section{The large space-based missions: K-EUSO and POEMMA}
\label{sec:keuso}
The central objective of K-EUSO~\cite{keuso} is the first consistent
measurement of the UHECR flux over the entire sky with almost uniform exposure.
K-EUSO is a result of the joint efforts to improve the performance of the
Russian KLYPVE mission~\cite{klypve}, by employing the technologies developed for the
JEM-EUSO mission, such as the optical system, focal surface detectors and the readout
electronics.
Since its first conception as KLYPVE, K-EUSO project has passed various
modifications aimed to satisfy carrier volume and deployment capabilities on the ISS.
It will be the first detector with a real capability for UHECR spectrum and
anisotropy study with a yearly exposure of
$\sim$2$\times$10$^4$ km$^2$ $yr$ $sr$ and a flat full celestial sphere
coverage. The adopted optical layout is a two-lens telescope of 240 $\times$ 120 cm$^2$ with a
FoV of 40 degree. The focal surface consists of about 40 PMDs.
The temporal (sampling time is 1$\mu$s) and spatial (angular resolution per
pixel 0.066$^{\circ}$) evolution of UV light recorded by K-EUSO will allow
the reconstruction of the EAS, with sufficient energy and arrival direction
resolutions above energies of 5$\times$10$^{19}$ eV~\cite{fenu_keuso_icrc21}.
Attached to the Russian MRM-1 module on-board ISS, K-EUSO is planned to operate
for minimum of 2 years and it can function more than 6 years if the lifetime of
 the ISS is extended.

The Probe Of Extreme Multi-Messenger Astrophysics (POEMMA)
mission~\cite{poemma} is being designed to establish charged particle astronomy from space
with UHECRs and cosmic neutrinos.
POEMMA will monitor colossal volumes of the Earth's atmosphere to
detect EASs produced by extremely energetic cosmic
messengers: cosmic neutrinos above 20 PeV and UHECRs above 20 EeV over the
entire sky.
The POEMMA design combines the concept developed for the OWL mission
~\cite{owl} and the experience of the JEM-EUSO fluorescence
detection camera. POEMMA is composed of two identical satellites flying in
formation at 525 km altitude with the ability to observe overlapping regions
during moonless nights at angles ranging from Nadir to just above the limb of
the Earth, but also with independent pointing strategies to exploit at maximum
the scientific program of the mission.
Each telescope is composed of a wide (45$^\circ$)) FoV Schmidt optical
system with an optical collecting area of over 6 m$^2$.
The POEMMA FS is composed of a hybrid of two types of cameras: about 90\%
of the FS is dedicated to the POEMMA fluorescence camera (PFC), while POEMMA
Cherenkov camera (PCC) occupies the crescent moon shaped edge of the
FS which images the limb of the Earth. The PFC is composed of JEM-EUSO
PDMs based on MAPMTs. The typical time between images for the PFC is about
1 $\mu$sec. The much faster POEMMA Cherenkov camera (PCC) is composed of
Silicon photo-multipliers (SiPMs) which will be tested
with EUSO-SPB2. The PFC registers
UHECR tracks from Nadir to just below the Earth's limb (above 20 EeV), while the PCC
registers light within the Cherenkov emission cone of up-going showers around
the limb of the Earth and also from high energy cosmic rays above the limb of
the Earth (above 20 PeV). Simulations indicate that POEMMA has excellent
performance in UHECR energy, $X_{max}$, and angular reconstruction capabilities~\cite{prd_poemma}, 
as well as high sensitivity in Target of Opportunity observations exploiting upward-moving EASs induced
by Earth-interacting tau neutrinos~\cite{too_poemma}. The roadmap to the POEMMA mission is summarized 
in~\cite{angela_icrc21}.

\section{Acknowledgments}
This work was partially supported by Basic Science Interdisciplinary Research Projects of
RIKEN and JSPS KAKENHI Grant (22340063, 23340081, and 24244042), by
the Italian Ministry of Foreign Affairs and International Cooperation,
by the Italian Space Agency through the ASI INFN agreements n. 2017-8-H.0 and n. 2021-8-HH.0,
by NASA award 11-APRA-0058, 16-APROBES16-0023, 17-APRA17-0066, NNX17AJ82G, NNX13AH54G, 80NSSC18K0246, 80NSSC18K0473, 80NSSC19K0626, and 80NSSC18K0464 in the USA,
by the French space agency CNES,
by the Deutsches Zentrum f\"ur Luft- und Raumfahrt,
the Helmholtz Alliance for Astroparticle Physics funded by the Initiative and Networking Fund
of the Helmholtz Association (Germany),
by Slovak Academy of Sciences MVTS JEM-EUSO, by National Science Centre in Poland grants 2017/27/B/ST9/02162 and
2020/37/B/ST9/01821,
by Deutsche Forschungsgemeinschaft (DFG, German Research Foundation) under GermanyÕs Excellence Strategy - EXC-2094-390783311,
by Mexican funding agencies PAPIIT-UNAM, CONACyT and the Mexican Space Agency (AEM),
as well as VEGA grant agency project 2/0132/17, and by by State Space Corporation ROSCOSMOS and the Interdisciplinary Scientific and Educational School of Moscow University "Fundamental and Applied Space Research".
\clearpage
\section*{Full Authors List: \Coll\ Collaboration}
%
%

\begin{sloppypar}
{\small \noindent
G.~Abdellaoui$^{ah}$,
S.~Abe$^{fq}$,
J.H.~Adams Jr.$^{pd}$,
D.~Allard$^{cb}$,
G.~Alonso$^{md}$,
L.~Anchordoqui$^{pe}$,
A.~Anzalone$^{eh,ed}$,
E.~Arnone$^{ek,el}$,
K.~Asano$^{fe}$,
R.~Attallah$^{ac}$
% ...
}
\end{sloppypar}
\vspace*{.3cm}

{ \footnotesize
\noindent
% ...
$^{ac}$ LPR at Dept. of Physics, Faculty of Sciences, University Badji Mokhtar, Annaba, Algeria \\
% ...
$^{ah}$ Telecom Lab., Faculty of Technology, University Abou Bekr Belkaid, Tlemcen, Algeria \\
% ...
$^{cb}$ Universit\'e de Paris, CNRS, AstroParticule et Cosmologie, F-75013 Paris, France\\
% ...
$^{ed}$ Istituto Nazionale di Fisica Nucleare - Sezione di Catania, Italy\\
% ...
$^{eh}$ INAF - Istituto di Astrofisica Spaziale e Fisica Cosmica di Palermo, Italy\\
% ...
$^{ek}$ Istituto Nazionale di Fisica Nucleare - Sezione di Torino, Italy\\
$^{el}$ Dipartimento di Fisica, Universita' di Torino, Italy\\
% ...
$^{fe}$ Institute for Cosmic Ray Research, University of Tokyo, Kashiwa, Japan\\
% ...
$^{fq}$ Nihon University Chiyoda, Tokyo, Japan\\
% ...
$^{md}$ Universidad Polit\'ecnia de madrid (UPM), Madrid, Spain\\
% ...
$^{pd}$ University of Alabama in Huntsville, Huntsville, AL; USA\\
$^{pe}$ Lehman College, City University of New York (CUNY), NY, USA\\
% ...
}

%

\end{document}